\documentstyle[12pt,epsf]{article}
\newif\ifall\alltrue
\def\private#1{}			% not print private "
\def\eqd{\stackrel{\bullet}{=}}

\def\gdw{\Longleftrightarrow}

\def\nq{\hspace{-1em}}

\def\ignore#1{}

\def\hbar{h\!\!\!\!^{-}\,}

\def\beq{\begin{equation}}
\def\eeq{\end{equation}}
\def\bqa{\begin{eqnarray}}
\def\eqa{\end{eqnarray}}

\begin{document}
%%%%%%%%%%%%%%%%%%%%%%%%%%%%%%%%%%%%%%%%%%%%%%%%%%%%%%%%%%%%%%%%%
%		       T i t l e - P a g e                    	%
%%%%%%%%%%%%%%%%%%%%%%%%%%%%%%%%%%%%%%%%%%%%%%%%%%%%%%%%%%%%%%%%%
\ifall

\hspace*{13cm}LMU 95-03

\hspace*{13cm}February 1995

\begin{center}					  \vspace*{3cm}
{\LARGE Gauge Invariant Quark Propagator         }\\[0.5cm]
{\LARGE  in the Instanton Background             }\\[3cm]
  {\bf Marcus Hutter\footnotemark}                \\[2cm]
  {\it Sektion Physik der Universit\"at M\"unchen}\\
  {\it Theoretische Physik}                       \\
  {\it Theresienstr. 37 $\quad$ 80333 M\"unchen} \\[2cm]
\end{center}
\footnotetext{E--Mail:hutter@hep.physik.uni--muenchen.de}

\begin{abstract}
After a general discussion on the choice of gauge, we compare
the quark propagator in the background of one instanton
in regular and singular gauge with a gauge invariant propagator
obtained by inserting a path-ordered gluon exponential.
Using a gauge motivated by this analysis, we were able to obtain
a finite result for the quark condensate without introducing
an infrared cutoff nor invoking some instanton model.

\end{abstract}
\newpage

\tableofcontents
%\listoffigures
%\listoftables
%\clearpage

\fi
\ifall
%%%%%%%%%%%%%%%%%%%%%%%%%%%%%%%%%%%%%%%%%%%%%%%%%%%%%%%%%%%%%%%%%
\section{Introduction}\label{sec1}
%%%%%%%%%%%%%%%%%%%%%%%%%%%%%%%%%%%%%%%%%%%%%%%%%%%%%%%%%%%%%%%%%
A variety of predictions concerning chiral symmetry breaking
and concerning the lightest hadrons in various channels can be
made within the instanton liquid model. Although there are various
attempts to derive this model from first principles, it is still
an open question, whether instantons melt or not. Thus the
infrared problem remains unsolved.

In section \ref{sec3}, I present a small calculation of the
quark condensate. I get a finite result by choosing an approriate
gauge and performing a self energy resummation.

Because the finiteness
essentially depends on the choice of gauge, I give a more general
discussion in section \ref{sec2} of how to choose a gauge when
calculating gauge dependent quantities. The quark propagator in
the background of one instanton in the well known regular and singular
gauge is compared to a gauge invariant (GI) propagator,
for which explicit expressions are calculated in this work.

The rest of section \ref{sec1} contains a summary of well known
instanton formulas, which are used in the subsequent sections.

%---------------------------------------------------------------%
\subsection{Instantons in QCD}
%---------------------------------------------------------------%
The solutions of the euclidian Yang-Mills equations fall into
topology classes $N\in Z\!\!\!Z$ and are called $N$
instanton\footnotemark\ solutions.
  \footnotetext{In this work instantons $(N>0)$ and anti-instantons
  $(N<0)$ will both be called instantons and are distinguished by
  their topological charge $Q=N$.}
The one instanton solution has the well-known form
\bqa\nonumber
  A_{I\mu}^a(x) &=& O_I^{ab}\eta_{b\mu\nu}^{Q_I}
                  { (x-z_I)_\nu \over (x-z_I)^2 }
                  { 2\rho^2 \over (x-z_I)^2+\rho^2 }
		  \quad\mbox{in singular gauge}  \\ \nonumber
  A_{I\mu}^a(x) &=& O_I^{ab}\eta_{b\mu\nu}^{-Q_I}
                  { 2(x-z_I)_\nu \over (x-z_I)^2+\rho^2 }
		  \quad\quad\quad\quad\quad\mbox{in regular gauge}
						 \\ \nonumber
\gamma_I=(z_I,O_I,\rho_I,Q_I)&=& {\mbox{(location, orientation, radius,
topological charge)}} \nonumber
\eqa
The parameters $\gamma_I$ of the instanton simply reflect the
symmetries of the Lagrangian (translation, rotation, scale invariance,
parity). The partition function in semiclassical approximation is
\cite{tHo},\cite{Ber}
\begin{eqnarray}
  Z_1 &=&
    {1\over 2}\sum_{Q_I=\pm 1} \int\!d^4\!z_I dO_I d\rho_I
    \, D(\rho_I) \;=\; V_4\int_0^\infty\nq d\rho\,D(\rho)
    \;=\; V_4 \bar D
  \nonumber\\
  D(\rho) &=& {C_{N_c}\over\rho^5}S_0^{2N_c}e^{-S_1(\rho) }
    \;=\; C_{N_c}\rho^{-5}S_0^{2N_c}(\rho\Lambda)^b
  \nonumber\\
    C_{N_c} &=& {4.6e^{-1.679N_c} \over\pi^2(N_c-1)!(N_c-2)! }
  \nonumber\\
  S_0 &=& {8\pi^2\over g_0^2} \quad,\quad
  S_1(\rho) \;=\; {8\pi^2\over g^2(\rho)} \;=\;
    b\ln{1\over\rho\Lambda} \quad,\quad
  \Lambda=\Lambda_{PV}
  \nonumber\\
  S_2(\rho) &=& {8\pi^2\over g^2(\rho)} \;=\;
    b\ln{1\over\rho\Lambda} + {b^\prime\over b}\ln\ln{1\over\rho\Lambda}
    + O({1\over\ln{1\over\rho\Lambda}}) \quad,\quad
  \nonumber
\end{eqnarray}
$D(\rho)$ is the density of instantons of size $\rho$,
$g(\rho)$ the running coupling constant, $b={11\over 3}N_c$ and
$b^\prime={17\over 3}N_c^2$.
$S_0$ is the classical instanton action and $g_0$ is the unrenormalized
tree level coupling constant. Whenever $g_0$ appears in some formulas,
one has to guess its value. This unlucky situation may be improved
by using the two loop expression for $D(\rho)$ replacing $S_0$ by $S_1$
and $S_1$ by $S_2$. But this is only an improvement for a small coupling.
When $\rho$ reaches the QCD scale $\Lambda$ one should
rely on a low order calculation to have a chance to get sensible
results.
%---------------------------------------------------------------%
\subsection{The Infrared Problem}
%---------------------------------------------------------------%
The sum of widely separated instantons is also an approximate
solution of the YM equations.
$$
  A=\sum_{I=1}^N A_I \quad,\quad S[A]\approx NS_0
$$
The partition function of this so called instanton gas is
$$
 Z=\sum_{N=0}^\infty Z_N \quad,\quad Z_N\approx {1\over N!}(V_4\bar D)^N
$$
The sum is dominated by an instanton
density $N/V_4=\bar D$. Unfortunately $\bar D$ is infinite and
the assumption of a dilute instanton gas is inconsistent.
This infinite density is caused by the divergence of $D(\rho)$
for large $\rho$, which in fact is a consequence of the increasing
coupling constant at large distances.

There were several suggestions to overcome this problem.
The most primitive is to introduce a cut-off $\rho_c$ and
ignore large instantons:
$$
  \bar D_{\rho_c}=\int_0^{\rho_c}\!\! d\rho\,D(\rho) \quad.
$$
The cut-off is chosen small enough to make the spacetime fraction $f$
filled with instantons less than 1 so that the dilute gas model
is justified
$$
  f = {2\over N_c}\int_0^{\rho_c}\!\! d\rho\,{1\over 2}\pi^2\rho^4 D(\rho)
      \quad < \quad 1
$$
This simple cut-off procedure can be improved by introducing
a scale invariant hardcore repulsion between instantons, which
effectively supresses large instantons \cite{Ilg}. This procedure has
the advantage of respecting the scaling Ward identities which
are otherwise violated by the simple cut-off ansatz.
In \cite{Dya} such an repulsion has been found
leading to a phenomenologically welcomed packing fraction.
Unfortunately this repulsion is an artefact of the sum-ansatz as
has been shown by \cite{Ver2}. Therefore the infrared problem is
still unsolved.

Nevertheless it is possible to make successful predictions
by simply assuming a certain instanton density and some average radius.
This instanton liquid model has been very successful in describing
the physics of light hadrons \cite{ShV},\cite{Hut2}.

In high energy processes involving momenta $p$ of $1-10$GeV, $D(\rho)$
is usually multiplied by a function sharply peaked at $\rho\sim p^{-1}$.
The integral over $\rho$ is now dominated by small instantons
and infrared convergent. The results are therefore independent of
the cut-off and no model has to be invented.

\fi
\ifall
%%%%%%%%%%%%%%%%%%%%%%%%%%%%%%%%%%%%%%%%%%%%%%%%%%%%%%%%%%%%%%%%%
\section{On the Choice of Gauge}\label{sec2}
%%%%%%%%%%%%%%%%%%%%%%%%%%%%%%%%%%%%%%%%%%%%%%%%%%%%%%%%%%%%%%%%%
%---------------------------------------------------------------%
\subsection{Generalities}
%---------------------------------------------------------------%
Gauge symmetry is a rather large symmetry, an infinite product
of $SU(N_c)$ in the case of QCD. A physicist is always happy
of having symmetries because they can be exploited to make
predictions even without solving the theory. Gauge symmetry is
neccessary to get a physical vector particle spectrum.
As long as one does not make an approximation which manifestly
breaks gauge symmetry one can choose a comfortable gauge for
calculations because the result is GI.
But it is very difficult {\it not} to break GI, especially
in a non-abelian gauge theory. It is not easy to find a GI
regularization and furthermore, the gluon propagator, the primary
object in perturbation theory, is not GI. Of course it is meanwhile
well known how to perform GI calculations in every order
perturbation theory using FP-ghosts and dimensional regularization.
Every new approach beyond perturbation theory is again confronted
with the problem of GI. In lattice theory the Wilson action had
to be invented. In Schwinger-Dyson and Bethe-Salpeter type
selfconsistency equations GI is still an open problem. In
instanton physics when going beyond the one instanton approximation
the choice of gauge is also important. This will be discussed in the
next paragraph. There is a related problem when considering
non-GI objects from the very beginning like the gluon or
quark propagator. Strictly
speaking they are only {\it defined} when relying to a certain gauge.
In principle one should not give them any physical meaning at all.
Often one is tempted to do so and therefore it is necessary
to give some motivation of choosing this or that gauge.
%---------------------------------------------------------------%
\subsection{A Natural Gauge}
%---------------------------------------------------------------%
The gauge field $A_\mu^a$ describes the connection between
neighbouring vector bundles over the spacetime manifold $I\!\!R^4$.
Thus a choice of gauge is like the choice of a coordinate system
in general relativity with connection $\Gamma^\mu_{\;\;\nu\rho}$.
When choosing a crooked coordinate system, although being in a smooth
universe, there will appear fictitious accelerations towering above
the real physical accelerations
$$
  \ddot x^\mu_{phys} = \ddot x^\mu_{fict} + \Gamma^\mu_{\;\;\nu\rho}
                       \dot x^\nu \dot x^\rho \quad.
$$
When making general covariant calculations
these fictitious accelerations and the $\Gamma$ contribution
will cancel out thus leading to the correct
small result. But the slightest unsystematic approximation
will produce gross errors. The natural solution of this problem is to
use a coordinate system as smooth as possible to avoid fictitious
accelerations, e.g. to choose $\Gamma^\mu_{\;\;\nu\rho}$ as small
as possible. To make this statement more quantitative we
may try to minimize $(\ddot x^\mu_{phys}-\ddot x^\mu_{fict})^2$
simultaneously for
all curves. This is done by choosing a coordinate system which
minimizes\footnote{In Euclidian space this is a positive definite norm}
$$
  ||\Gamma||^2 \;:=\; \int \Gamma_\mu^{\;\;\nu\rho}
                           \Gamma^\mu_{\;\;\nu\rho}
                           \,d^4\!x
$$
This obviously measures the crookedness of the coordinate system.

Let us now transfere this to QCD. The analog norm for the
gauge potential is
$$
  ||A||^2 \;:=\; \int tr_c A_\mu A^\mu \,d^4\!x
$$
A stationary point is found by variating $||A||$ w.r.t.
gauge transformations
$$
  \delta A_\mu = i[A_\mu,\Omega]+\partial_\mu\Omega \quad,\quad
  \delta ||A||^2 = 2i\int tr(\partial_\mu A^\mu)\Omega d^4\!x = 0
  \quad \forall\Omega   \quad\gdw\quad
  \partial_\mu A^\mu =0
$$
Therefore in Lorentz gauge, $A_\mu^a$ contains as few pure gauge
as possible, if the stationary point is a minimum.
An expansion in $A$ is thus most rapidly convergent in Lorentz gauge.
\private{Im Pfadintegral Feynman gauge}
In applications where $A$ is not needed in total e.g. when
only a certain momentum region is probed, different norms and
different gauges may be optimal in the sense discussed above.
Especially one should include derivatives of $A$ into the
norm in order to guarantee a smooth $A$ which is important for
high energies.
%---------------------------------------------------------------%
\subsection{On the Gauge in Instanton Physics}
%---------------------------------------------------------------%
When calculating GI quantities in the background of
one instanton in a GI way the choice of gauge is only a matter
of convenience. But one can see that there are large cancelations
between different terms in regular gauge at large distances due
to their slow decay and in singular gauge at small distances
due to the topological singularity at the instanton center.
For non-GI invariant quantities like the gluon or quark propagator,
or when making some unsystematic approximation,
the lesson is to use singular/regular gauge when dealing with
low/high energies to avoid these cancelations.
This is consistent with the discussion
given above. Singular as well as regular gauge fulfill
the Lorentz condition. $||A_{sing}||$ is finite and a minimum.
$A_{sing}$ is therefore a good choice for low energies.
For high energies it is important to have a smooth $A$ which
is obviously only satisfied by the regular gauge.

To linearly superpone instantons they have to decay
rapidly enough. Therefore one has to use singular gauge.
This argument can in principle be circumvented
by superponing two fields $A_N$ and $A_{\overline{N}}$ the
former/latter being an exact multi-instanton/anti-instanton
configuration in regular gauge.
Despite this, for low energies singular gauge is in any case
a good choice and for high energies a one instanton approximation
is already a good approximation.
%---------------------------------------------------------------%
\subsection{The Quark Propagator in Axial Gauge}
%---------------------------------------------------------------%
A specific example to test the gauge dependence is the quark
propagator. The contribution of one instanton of radius $\rho$
to $M(p)=ip^2\bar S_I(p)$
which usually is interpreted as a constituent quark mass is shown
in figure \ref{fig1} in regular, singular and axial gauge.
The regular graph is larger than the singular at low momentum
and the singular graph shows the slow decay (only polynomial in $1/p$)
for large momenta. The analytical expressions are well known
and are listed in appendix \ref{appa} together with expressions
in axial gauge which will be derived and discussed below.

A correlator containing color-non-singlet operators can be made GI
by connecting distant points with a special path-ordered exponential
containing the gauge field.
The exponential ensures the parallel transport of color
from one point to the other.
The GI quark propagator may symbolically written as
\beq\label{gi31}
  S_{ax}(x,y) = \langle 0|\Psi(x)
           P\exp\left(i\int_x^y\!\! dz\!\cdot\!A(z)\right)
           \bar\Psi(y)|0\rangle
\eeq
$P$ denotes path-ordering.
We have already defined $S_{ax}$ to be its color singlet part
because only the singlet part is GI.
$S_{ax}$ will be called the axial propagator because in axial gauge
with $n_\mu=x_\mu-y_\mu$ the exponential vanishes.
In the one instanton background in zeromode approximation we get
$$
  S_{ax}(x,y) = {1\!\!1_c\over N_c} tr_c\left[
           P\exp\left(i\int_x^y\!\! dz\!\cdot\!A(z)\right)
           \psi(x)\bar\psi(y) \right]
$$
where $A$ is now the instanton field and $\psi$
is the zeromode in any gauge.
In a coordinate system where the instanton sits at the origin and
$x-y$ is in time direction\footnote{
Although working in Euclidian space we will adopt the Minkowskian
language $(x_0, {\bf x}) =$ (time, space).}
$({\bf x}={\bf y}={\bf z})$ the path ordered exponential reduces
to an ordinary exponential. Alternatively we could have tried
to find a gauge transformation which transforms the regular gauge
in axial gauge. In both cases we get:
\bqa\label{gi33}
  S_{ax}(x,y) &=& {1\over N_c}tr_c\left[\psi_{ax}(x)
                \bar\psi_{ax}(y)\right] \quad,\quad
  \psi_{ax}(x) = R(x)\psi_{reg}(x) ,
\\ \label{gi34}
  R(x) &=& e^{\pm i\alpha(x){{\bf\tau}\!\cdot\!{\bf x} \over |{\bf x}| } }
       = \cos\alpha(x)\pm i{{\bf\tau}\!\cdot\!{\bf x} \over |{\bf x}| }
         \sin\alpha(x) =: \pm i\tau_\mu^\pm\!\cdot\! \tilde{x}(x)
\\ \label{gi35}
  \alpha(x) &=& {|{\bf x}|\over\sqrt{{\bf x}^2+\rho^2}}
              \arctan{x_0\over\sqrt{{\bf x}^2+\rho^2}}
\eqa
$\alpha(x)$ may also be written in a covariant form
$$
  \alpha(x)=\pm\left(1+{\rho^2(x-y)^2\over x^2y^2-(xy)^2}\right)^{-1/2}
  \arctan\sqrt{(x^2-(xy))^2\over x^2y^2-(xy)^2+\rho^2(x-y)^2}
$$
but now $\alpha(x)$ depends also on $y$ and the expression for the
propagator no longer factorizes. The reason for this is that the
axial gauge is not covariant, but the definition of the
propagator is. Inserting (\ref{gi34}) and (\ref{gi86}) into
(\ref{gi33}) we get
\beq\label{gi36}
  S_{ax}(x,y) = {1\over N_c} \left( (\tilde{x}\tilde{y})
     - {1\over 2}\tilde{x}_\mu\tilde{y}_\nu\sigma^{\mu\nu} \right)
       {1\pm\gamma_5\over 2}\varphi_{reg}(x)\varphi_{reg}(y)
\eeq
Inserting (\ref{gi34}), (\ref{gi35}) and (\ref{gi86})
into (\ref{gi36}) the space-time averaged propagator can
be expressed as an integral over elementary functions
$$
  \bar S(x-y) = \int_0^\infty\nq dr\int_{-\infty}^\infty\nq dt\,4\pi r^2
     \cos\left[{r\over R}
       \left(\arctan{t+|x-y|\over R} -
             \arctan{t\over R}
       \right) \right]\cdot
$$
$$
  \cdot{1\over 2N_c}{\rho^2\over
     \pi^2(R^2+(t+|x-y|)^2)^{3/2} (R^2+t^2)^{3/2} }
  \quad,\quad R^2 = r^2+\rho^2
$$
The difference between the propagator in regular and axial gauge
is the insertion of the $\cos[\ldots]$ factor. Therefore the
axial propagator is everywhere smaller than the regular propagator,
except at $x=y$ where they coincide because the path-ordered
exponential is one.
At large distances it is smaller by a factor $\pi/4$.
Instead of performing the integration in coordinate space, let us
go directly to the more interesting momentum space representation:
$$
  \bar S_I(p) = {1\over 2N_c}\varphi_{ax}^\mu(p)
    \varphi_{ax}^{\mu\dagger}(p)  \quad,\quad
  \varphi_{ax}^\mu(p) = \int \tilde{x}^\mu(x)\varphi_{reg}(x)e^{ipx}dx
$$
Although $\varphi_{ax}^\mu$ does not transform like a vector,
we can choose a convenient direction of $p$ because
$\varphi\varphi^\dagger$ is a Lorentz scalar. For pure spacelike $p$
the spacial components of $\varphi$ vanish because the integrand
is anti-symmetric w.r.t time reflection. Only the time component is
nontrivial
$$
  \varphi_{ax}^0(p) = \int\! d^3\!r\int\!dt\,
    \cos\left[{r\over R}
       \arctan{t\over R} \right]
  {\rho\over\pi(R^2+t^2)^{3/2}}
  e^{i{\bf p}\cdot{\bf r}}
$$
With the following hints
\bqa
  \cos(\gamma\arctan x) &=& \mbox{Re}
    \left({1+ix\over 1-ix}\right)^{\gamma/2}		   \nonumber\\
  \int_{-\infty}^\infty\!\! (R-it)^{-\alpha}(R+it)^{-\beta} dt &=&
    2\pi(2R)^{1-\alpha-\beta}
  {\Gamma(\alpha+\beta-1)\over\Gamma(\alpha)\Gamma(\beta)} \nonumber\\
    \Gamma({3\over 2}-x)\Gamma({3\over 2}+x) &=&
  {(1/4-x^2)\pi\over\cos\pi x}                             \nonumber\\
    \int d^3\!r\,e^{i{\bf p}\cdot{\bf r}}f(r) &=&
    {2\pi\over p}\int_0^\infty\!\! f(r)\sin(pr)r\,dr
\eqa
the reader should be able to perform the $t$
and the angular integration $d\Omega_r$,
$$
  \varphi_{ax}^0(p) = {8\over p\rho}\int_0^\infty\nq
  \cos\left({\pi r\over 2R}\right)\sin(pr)r\,dr
$$
I was not able to perform this last integral analytically,
but for small momenta it is easy to see that $\varphi_{ax}^0(p)$
behaves like $\pi^2\rho/p$. For large $p$ it decays like
$\sim e^{-p\rho}$ with a non-polynomial coefficient because
of an essential singularity at $r=\pm i\rho$.
Comparing $\varphi_{reg}$, $\varphi_{sing}$ and $\varphi_{ax}$ plotted
in figure \ref{fig1} we see that the axial $\varphi$ lies somewhat
in between the regular and the singular. So one may conclude
that axial gauge is a good compromise for all momenta.

The calculation of the GI propagator seems to make the discussion
of its gauge dependent partners obsolet. I will now argue that
this is not the case. The reason is that there are a huge
number of GI definitions of a quark propagator and (\ref{gi31})
is only one
possible choice. One obvious generalization is to choose a more
complicated path from $x$ to $y$ than a straight line.
The next thing one could do is not to restrict oneself to a specific
path, but to take into account all paths one is interested in
and average the results with arbitrary weights. Another
possibility is to let the path depend on the gauge
field itself, as long as this choice is made in a GI way.
Finally one can combine both generalizations. I am sure that
it is possible to produce any result for the propagator with
a suitable generalized definition.
The advantage of the standard axial propagator is, that the
definition is simple and that
the non local operator has a physical interpretation. It creates
a quark-antiquark pair connected by a thin gluon flux tube.
This might be a good choice for a non-local meson creation operator.
But it is also plausible that one of the generalizations given
above is even better. The only thing I want to point out is,
that the GI definition for the propagator given above is
nothing more than to work in axial gauge.
One still has to choose the right gauge using more sophisticated
arguments.
\fi
\ifall
%%%%%%%%%%%%%%%%%%%%%%%%%%%%%%%%%%%%%%%%%%%%%%%%%%%%%%%%%%%%%%%%%
\section{The Quark Condensate}\label{sec3}
%%%%%%%%%%%%%%%%%%%%%%%%%%%%%%%%%%%%%%%%%%%%%%%%%%%%%%%%%%%%%%%%%
%---------------------------------------------------------------%
\subsection{$N_c\to\infty$}
%---------------------------------------------------------------%
The only reason for performing the $N_c\to\infty$ limit is
to make the $N_c$ dependence of the resulting formulas simple.
The accuracy has been checked to be within the standard 10\%
for $N_c=3$ usually achieved by $1/N_c$ expansion.
Here the accuracy can simply be understood. The actual expansion
parameter is not $1/N_c=1/3$ itself but $1/b\approx 1/11$.
The following asymptotic formulas will be used
\bqa
  N_c!^{1/N_c} \eqd N_c/e \quad,\quad b \eqd {11\over 3}N_c \quad,\quad
  C_{N_c}^{1/b} \eqd 2.22b^{-6/11} \,	 \nonumber\\
  \rho^5 D(\rho) \sim (2.22(S_0/b)^{6/11}\rho\Lambda)^b \quad,\quad
  S_0/b = {24\pi^2\over 11}(g_0^2 N_c)^{-1}
\eqa
Every equality in the large $N_c$ limit will be marked with a dot.
Notice that in this limit instantons of size
$\rho < {1\over 2.22}(S_0/b)^{-6/11}\Lambda^{-1}$ are completely
suppressed. Above this threshold the instanton density gets
infinite. $S_0/b$ is independent of $N_c$ because
$g_0\sim 1/\sqrt{ N_c}$.

%---------------------------------------------------------------%
\subsection{Effective Quark Mass}
%---------------------------------------------------------------%
In the presense of one light quark flavor the instanton density
$D(\rho)$ has to be multiplied with the functional determinant of the
Dirac operator
$$
  Det(iD\!\!\!\!/+im) \approx 1.34m\rho
$$
which is proportional to $m$ because of a zeromode of $D\!\!\!\!/$.
The quark propagator in the background of one instanton
is dominated by this zeromode
$$
  S_I(p,q)={\psi_I(p)\psi^\dagger_I(q) \over im}
$$
Averaging this expression over all collective coordinates $\gamma_I$
one gets
$$
  M(p) := ip^2\bar S_I(p) = {1.34\over{2N_c}}\int_0^\infty\nq d\rho\,
          p^2\rho D(\rho) \varphi^2(p)
$$
Summing the contribution to the propagator
of $0,1,2,3,\ldots$ instantons, which is the analog of a selfenergy
resummation in perturbation theory,
$$
  S(p) = {1\over p\!\!/} +
         {1\over p\!\!/}{M(p)\over i}{1\over p\!\!/} +
         {1\over p\!\!/}{M(p)\over i}{1\over p\!\!/}
         {M(p)\over i}{1\over p\!\!/} + \ldots
       = {1\over p\!\!/+iM(p)}
$$
justifies to call $M(p)$ a dynamical quark mass.
Expressions of $\varphi$ in various gauges are given in
appendix \ref{appa}.
The graphs of $\varphi(p)$ in singular
and regular gauge cross over at
$$
  \varphi_{sing}(p)=\varphi_{reg}(p) \quad\gdw\quad
  2p\rho\approx 2.5
$$
Therefore one should use regular gauge for large $\rho$ and
singular gauge for small $\rho$. This choice of gauge also
makes the integral convergent for large $\rho$. At this stage we
have no infrared problem.
Using regular gauge in the whole integration interval we get
$$
  M_{reg}(p) = Bp({\Lambda\over 2p})^b \quad,\quad
  B=1.34\cdot16\pi^2(C_{N_c}b^{2N_c})(S_0/b)^{6b/11}I_b/N_c
$$
$$
  I_b = \int_0^\infty\nq dz\,z^{b-2}e^{-z} = (b-2)! \quad,\quad
  z=2p\rho
$$
The integral is sharply dominated by
$$
  z \eqd b\pm b^{1/2} >> 2.5
$$
therefore the result is independent of the choice of gauge for
$z<2.5$ justifying our use of regular gauge over the whole
integration interval. Axial gauge would lead to nearly the
same result as can be seen from figure \ref{fig1}.
Using singular gauge for large
$\rho$ would produce a divergent integral dominated by arbitrary
large instantons  inconsistent with the choice of gauge discussed
above.
The infrared "problem" shows up in the rapid raise of $M(p)$
for low $p$, which effectively supresses the propagation
of quarks with low virtuality $p^2$.
Consider some process involving quarks at distances $x=1/p_c$.
The effective quarkmass $M(1/x)$ is dominated by instantons
of much larger size
$$
  \rho=\rho_c(1\pm b^{-1/2}) >> x \quad,\quad \rho_c={b\over 2p_c}.
$$
In other words, given
an instanton of radius $\rho$ influences the physics at a much
smaller scale $x={2\over b}\rho << \rho$. Therefore the
interior of the instanton is probed and one should avoid
the singularity at its center by using regular gauge.

%---------------------------------------------------------------%
\subsection{The Quark Condensate}
%---------------------------------------------------------------%
Let us now calculate a real physical gauge invariant observable,
the quark condensate
$$
  \langle \bar\psi\psi\rangle :=
  \lim_{x\to 0}\mbox{tr}_{CD}(S(x)-S_0(x)) =
  -4iN_c\int {M(p)\over p^2+M^2(p)} {d^4\!p\over(2\pi)^4}
%  N_c\int\!\dbar^4\!p\mbox{tr}_D(S(p)-S_0(p))
$$
Inserting M(p) and performing the angular integration we get
$$
  |\langle \bar\psi\psi\rangle| =
  {N_c\over 16\pi^2}B^{3/b}J_b\Lambda^3
$$
$$
  J_b = \int_0^\infty {z^{b+2}\over 1+z^{2b}} dz =
  {\pi\over 2b\sin({b+3\over 2b}\pi)} \eqd {\pi\over 2b}
  \quad,\quad p=B^{1/b}{\Lambda\over 2}z
$$
The integral is finite and sharply dominated by $z\eqd 1\pm b^{-1}$.
Without resummation of the selfenergies the integral $J_b$
and thus condensate
would have turned out to be infinite.
The condensate is dominated by quark wavefunctions with
momenta
$$
  p=p_c(1\pm b^{-1}) \quad,\quad p_c=\beta b{\Lambda\over 2}
  \quad,\quad
  \beta := {1\over b}B^{1/b} \eqd {2.22\over e}(S_0/b)^{6/11}
$$
and depends on $\Lambda$ and $g_0$
$$
  |\langle \bar\psi\psi\rangle|^{1/3} =
  0.139\beta b\Lambda \quad.
$$
%---------------------------------------------------------------%
\subsection{Discussion}
%---------------------------------------------------------------%
Expressing $p_c$ and $\rho_c$ in terms of
$|\langle\bar\psi\psi\rangle|$ by eliminating $\beta$
we get our main result
\bqa
  p_c    &=& 3.59 |\langle \bar\psi\psi\rangle|^{1/3}  \nonumber \\
  \rho_c &=& {1.96\over N_c} |\langle \bar\psi\psi\rangle|^{1/3} \\
  2.22N_c\Lambda &=& (g_0^2N_c)^{6/11}|\langle\bar\psi\psi\rangle|^{1/3}
\eqa
A weak point is the experimental extraction of $g_0$.
It should be extracted from a reliable treelevel process at low
energies presumably of the order of $\rho_c$.
In QCD improved Bag-Models the main nonperturbative effect is
modeled by the bag and the hyperfinesplitting is caused
by a one gluon exchange. $g_0$ extracted from $\Delta-N$ splitting
is \cite{Clo}
$$
g_0^{bag} \approx 2.6
$$
Let me also give a theoretical guess of $g_0$.
The change to a two loop expression for the instanton density
$S_{0/1} \leadsto S_{1/2}$ can be effectively performed
by only replacing $S_0$ in the following way
$$
  (S_0/b)^{6/11} \leadsto (\ln{1\over \rho\Lambda})^\alpha
  \quad,\quad \alpha = {15\over 121}
$$
Because $\alpha$ is very small
$(\ln{1\over\rho\Lambda})^\alpha$ is approximately one
in a large range of values for $\rho\Lambda$.
and for
$$
  g_0^{guess} = 2.7\sqrt{3/N_c}
$$
the two 2 loop density coincides with the one loop density.
Because we do not believe that the 2 loop density is an improvement,
one should not take $g_0^{guess}$ too seriously.
At least it is not in contradiction with $g_0^{bag}$.

The condensate is well known to be
$|\langle \bar\psi\psi\rangle|^{1/3}=240\mbox{MeV}$.
Setting $N_c=3$ and taking $g_0=2.6$ for granted we get
\bqa
  p_c   \pm\Delta p   &=& (860\pm 80)\mbox{MeV}      \nonumber\\
  \rho_c\pm\Delta\rho &=& (160\pm 50)^{-1}\mbox{MeV}          \\
  \Lambda_{PV}  &\approx& 190\mbox{MeV}              \nonumber
\eqa
The most interesting thing is, that the condensate is sharply
dominated by quark field wave functions of rahter large
momentum $p_c$. On the other hand the dominating instantons
have a very large radius $\rho_c$, 4 times larger than usually
assumed in instanton liquid models. Nevertheless the predicted value of
$\Lambda_{PV}$, which of course must be assigned a large error
because of the rough estimate of $g_0$, is in agreement with
experiment.

\fi
%%%%%%%%%%%%%%%%%%%%%%%%%%%%%%%%%%%%%%%%%%%%%%%%%%%%%%%%%%%%%%%%%
\section{Conclusion}\label{sec4}
%%%%%%%%%%%%%%%%%%%%%%%%%%%%%%%%%%%%%%%%%%%%%%%%%%%%%%%%%%%%%%%%%
\ifall

Whenever one is calculating gauge dependent objects or when making
gauge breaking approximations, one is confronted with the problem
of choosing a "good" gauge. Specializing the general discussion
of section \ref{sec2} to the case of instantons, we came to the
conclusion that the regular gauge is appropriate for small distances
and the singular gauge for processes involving large distances.
The GI propagator was defined, calculated and compared to
the propagator in singular and regular gauge (figure \ref{fig1}).
The conclusion was,
that the GI propagator is not a-priori a good choice,
but lies somewhat in between regular and singular gauge.

Using an appropriate gauge along the lines discussed in section
\ref{sec2} we were able to derive a finite quark condensate
without taking an infrared cut-off for the instanton radius nor
relying on some instanton model. The linear relation between
$|\langle\bar\psi\psi\rangle|^{1/3}$ and the QCD scale $\Lambda$
is in agreement with experiment. The condensate if formed by
quark fields of high momenta $p_c=860$MeV mainly lying within
the sharp region $\Delta p=80$MeV. The dominating instantons
are very large ($\rho_c=160$MeV).

{\bf Acknowledgements}
I want to thank the "Deutsche Forschungs-Gemeinschaft" for
supporting this work.

\fi
\begin{appendix}
\ifall
%%%%%%%%%%%%%%%%%%%%%%%%%%%%%%%%%%%%%%%%%%%%%%%%%%%%%%%%%%%%%%%%%
\section{Instantons in Singular, Regular and Axial Gauge}\label{appa}
%%%%%%%%%%%%%%%%%%%%%%%%%%%%%%%%%%%%%%%%%%%%%%%%%%%%%%%%%%%%%%%%%
The instanton at the origin in standard orientation is
given in singular, regular and axial gauge:
\bqa
   A_\mu^{sing}(x) &=& \eta^\pm_{\mu\nu} {x_\nu\over x^2}
     {\rho^2\over x^2+\rho^2} \quad\quad,\quad\quad
     \tau_\mu^\pm\tau_\nu^\mp=\delta_{\mu\nu}+i\eta_{\mu\nu}^\pm
                                                        \nonumber\\
   A_\mu^{reg}(x)  &=& \eta^\mp_{\mu\nu}
     {x_\nu\over x^2+\rho^2} \quad\quad\quad,\quad\quad
     \tau_\mu^\pm = (\pm i,{\bf\tau})    		\nonumber\\
   A_\mu^{ax}(x)   &=& R(x)A_\mu^{reg}(x)R^\dagger(x)
     + iR(x)\partial_\mu R^\dagger(x)
\eqa
The upper/lower sign corresponds to an instanton/anti-instanton
($Q=\pm 1$).
$$
  R(x) = \pm i\tau_\mu^\pm\tilde{x}^\mu(x) \quad,\quad
  \tilde{x}^\mu(x) = { \cos\alpha(x) \choose
    {{\bf x}\over |{\bf x}|}\sin\alpha(x) }
$$
$$
  \alpha(x) = {|{\bf x}|\over\sqrt{{\bf x}^2+\rho^2}}
              \arctan{x_0\over\sqrt{{\bf x}^2+\rho^2}}
$$
The covariant derivative $D\!\!\!\!/$ has one zeromode
$$
  iD\!\!\!\!/\psi = (i\partial\!\!\!/-A\!\!\!/)\psi = 0
$$
where the zeromode has the following form:
\bqa
  \psi_{sing}(x)   &=& \sqrt{2}\varphi_{sing}(x)x\!\!/\chi \quad,\quad
  \varphi_{sing}(x) = {\rho\over\pi|x|(x^2+\rho^2)^{3/2}} \nonumber\\
  \label{gi86}
  \psi_{reg}(x)    &=&\sqrt{2}\varphi_{reg}(x)\chi \quad\quad,\quad
  \varphi_{reg}(x) = {\rho\over\pi(x^2+\rho^2)^{3/2}}              \\
  \psi_{ax}(x)     &=& \sqrt{2}\varphi_{reg}R(x)\chi       \nonumber
\eqa
$\chi$ is a color Dirac spinor given by
$$
  \chi^\pm\bar\chi^\pm = {1\over 16}\gamma_\mu\gamma_\nu
    {1\pm\gamma_5\over 2}\tau^\mp_\mu\tau^\pm_\nu
$$
For light quarks the propagator is dominated by the zeromode.
When averaged over the instanton orientation, position and charge
the propagator is diagonal in momentum space and given by
$$
   \langle\psi(p)\psi^\dagger(p)\rangle =
   {1\over 2N_c}\varphi^2(p)
$$
where
\bqa
  \varphi_{sing}(p) &=& \pi\rho^2{d\over dz}
    [I_1(z)K_1(z)-I_0(z)K_0(z)]_{z=p\rho/2}
% = \left\{
%  \begin{array}{c@{\quad}c}
%    {2\pi\rho\over  p}     & p\rho\ll 1 \\
%    {12\pi\over p^4\rho^2} & p\rho\gg 1
%  \end{array} \right.
							\nonumber\\
  \varphi_{reg}(p)  &=& {4\pi\rho\over p}e^{-p\rho}	\nonumber\\
  \varphi_{ax}(p)   &=& {8\over p\rho}\int_0^\infty\nq
  \cos\left({\pi r\over 2\sqrt{r^2+\rho^2}}\right)\sin(pr)r\,dr
% = \left\{
%  \begin{array}{c@{\quad}c}
%    {\pi^2\rho\over  p}     & p\rho\ll 1 \\
%    {\sim e^{-p\rho}}       & p\rho\gg 1
%  \end{array} \right.
\eqa
The asymptotics are given in the following table
\begin{table}[hhh]\label{tab3}
  \begin{center}\begin{tabular}{|c|c|c|c|}               \hline
    ${p\over\rho}\varphi(p)$ & singular & regular & axial   \\ \hline
    $p\rho\ll 1$             & $2\pi$   & $4\pi$  & $\pi^2$ \\ \hline
    $p\rho\gg 1$             & ${12\pi\over(p\rho)^3}$
      & $4\pi e^{-p\rho}$    & $\sim e^{-p\rho} $        \\ \hline
\end{tabular}\end{center}
\vspace{-3ex}
\caption{
  \it Asymptotic behaviour of ${p\over\rho}\varphi(p)$
}\end{table}

The constituent mass of a quark in the gas approximation is

$$
  M(p) = ip^2\bar S(p) = 1.34\int_0^\infty\nq d\rho\,\rho D(\rho)
   \langle\psi(p)\psi^\dagger(p)\rangle
$$
Only when the instanton radius is kept fixed, the mass
is proportional to
$$
  M(p) \sim p^2\varphi^2(p) \quad.
$$
$p^2\varphi^2(p)$ is plotted in figure \ref{fig1}
in all three gauges.

\fi
%%%%%%%%%%%%%%%%%%%%%%%%%%%%%%%%%%%%%%%%%%%%%%%%%%%%%%%%%%%%%%%%%
%                  B i b l i o g r a p h y                    	%
%%%%%%%%%%%%%%%%%%%%%%%%%%%%%%%%%%%%%%%%%%%%%%%%%%%%%%%%%%%%%%%%%
\ifall
\stepcounter{section}
\addcontentsline{toc}{section}{\Alph{section} References}
\parskip=0ex plus 1ex minus 1ex

\fi
%\clearpage
%%%%%%%%%%%%%%%%%%%%%%%%%%%%%%%%%%%%%%%%%%%%%%%%%%%%%%%%%%%%%%%
\section{Figures}\label{Chf}
%%%%%%%%%%%%%%%%%%%%%%%%%%%%%%%%%%%%%%%%%%%%%%%%%%%%%%%%%%%%%%%
\ifall
\begin{figure}
   \epsfbox[85 0 480 550]{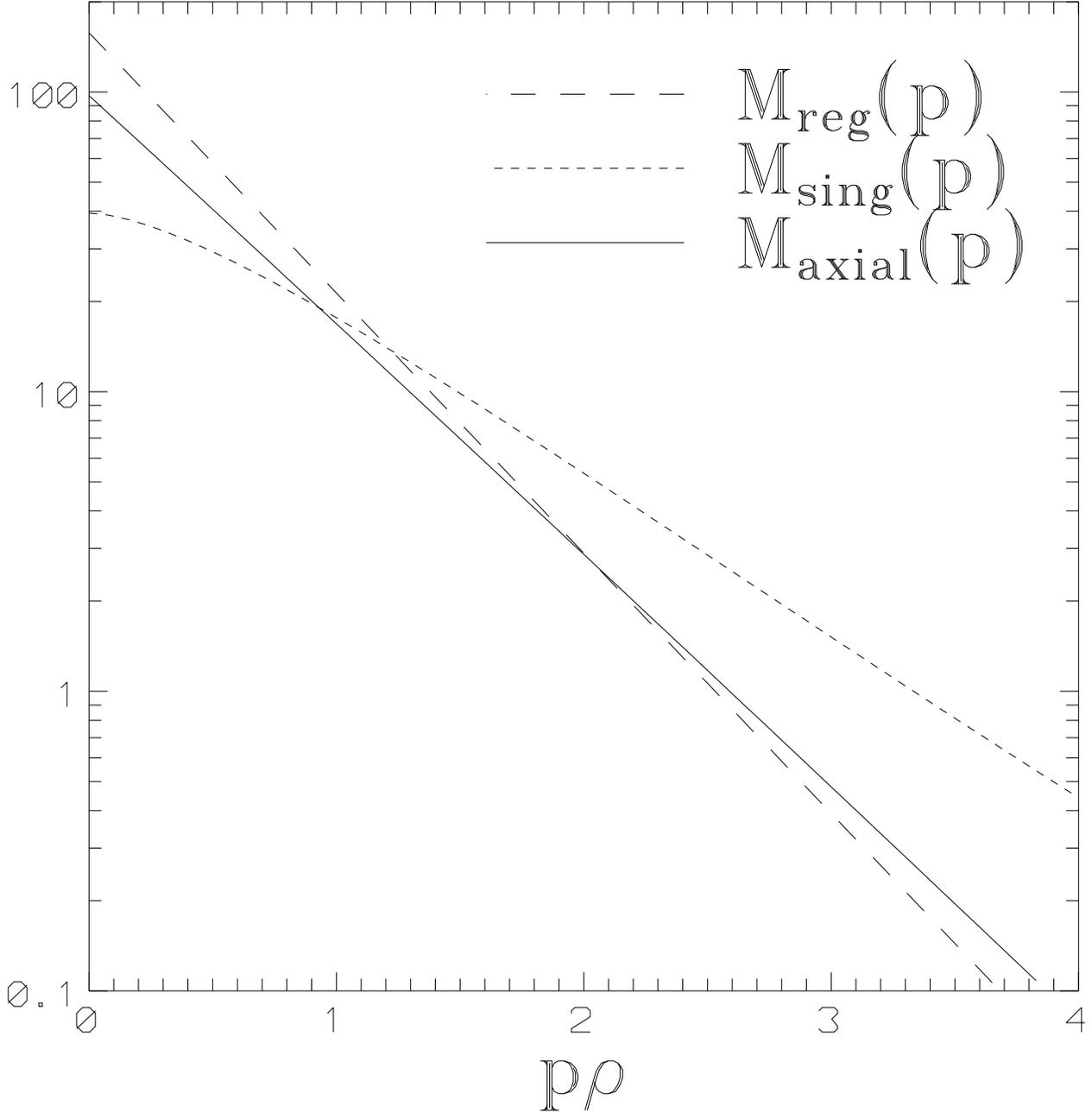}
   \caption{\label{fig1}
            \it Constituent quark mass $M(p)\sim p^2\varphi^2(p)$
            in singular, regular
            and axial gauge for fixed instanton radius $\rho$
            in arbitrary normalization. For a given momentum
            the corresponding lowest curve may be interpreted as
            the "most physical" one.}
\end{figure}
\fi
%%%%%%%%%%%%%%%%%%%%%%%%%%%%%%%%%%%%%%%%%%%%%%%%%%%%%%%%%%%%%%%%%
\end{appendix}
\end{document}